\documentclass{mem}
\usepackage{natbib}\usepackage{txfonts}\usepackage{balance}
\usepackage{graphicx}
\usepackage[a4paper,breaklinks,dvipdfm]{hyperref}
\idline{75}{282}
\begin{document}
\def\teff{$T\rm_{eff }$}
\def\kms{$\mathrm {km s}^{-1}$}

\title{Gaia Detection Capabilities of Spectroscopic Brown Dwarf Binaries}

   \subtitle{}

\author{
V. \,Joergens\inst{1,2}
\and S. \,Reffert \inst{3}}


\institute{
 Max-Planck Institut f\"ur Astronomie, 
             K\"onigstuhl~17, 69117 Heidelberg, Germany,\\
             \email{viki@mpia.de}
	\and  
	Institut f\"ur Theoretische Astrophysik,
        Zentrum f\"ur Astronomie der Universit\"at Heidelberg,
	Albert-Ueberle-Str. 2,
	69120 Heidelberg, Germany
        \and
        Landessternwarte, Zentrum f\"ur Astronomie der Universit\"at Heidelberg,
        K\"onigstuhl~12, 69117 Heidelberg, Germany
}

\authorrunning{Joergens et al.}

\titlerunning{Gaia Detection Capabilities of Spectroscopic Brown Dwarf Binaries}

\abstract{
  {The astrometric space mission Gaia is expected to detect a large number of brown dwarf binary
systems with close orbits
and determine astrometric orbit solutions. This will provide key information 
for the formation and evolution of brown dwarfs, such as the binary frequency and 
dynamical masses. 
Known brown dwarf binaries with orbit constraints from other techniques will
play an important role.
We are carrying out one of the most precise and long-lasting radial velocity surveys
for brown dwarf binaries in the Cha\,I star-forming region at the VLT. 
We were able to add two orbit determinations to the very small group of brown dwarf 
and very low-mass binaries with characterized RV orbits. We show here that the astrometric 
motion of both systems can be 
detected with Gaia. We predict an astrometric signal of 
about $1.2-1.6$ milliarcseconds (mas) for the brown dwarf binary Cha\,H$\alpha$\,8 and 
of $0.4-0.8$\,mas for the very low-mass binary CHXR\,74. 
We take the luminosity of the companions into account for these estimates
and present a relation for the astrometric signature of 
a companion with non-negligible luminosity.
}

  \keywords{brown dwarfs - stars: pre-main sequence -
planetary systems - astrometry - binaries: spectroscopic -
	     stars: individual (Cha\,H$\alpha$\,8, CHXR\,74)
               }
}
\maketitle{}

\section{Introduction}
\label{sect:intro}

Brown dwarf binaries play an important role in stellar astrophysics,
as they are key objects to constrain formation and evolution in the substellar regime.
Not only are the frequency and properties of brown dwarf binaries 
a pre-requisite to understand brown dwarf formation,
but brown dwarf binaries also provide - apart from microlensing (e.g., Gould et al. 2009) -
the only means to determine absolute masses from their orbits. 
As the mass is the most fundamental parameter for the evolution of a (sub)stellar object,
mass measurements, in particular at young ages (e.g., Stassun et al. 2006), 
are invaluable to establish the initial mass function, 
and to test and constrain evolutionary and atmospheric models.
Furthermore, the search for low-mass companions opens the
possibility to detect even lower mass and cooler objects (e.g., Joergens \& M\"uller 2007).
It addresses the question of the existence and frequency of planets around brown dwarfs,
which is an important empirical constraint for planet formation models. 

\begin{figure}[t]
  \centering
\resizebox{\hsize}{!}{\includegraphics[clip=true]{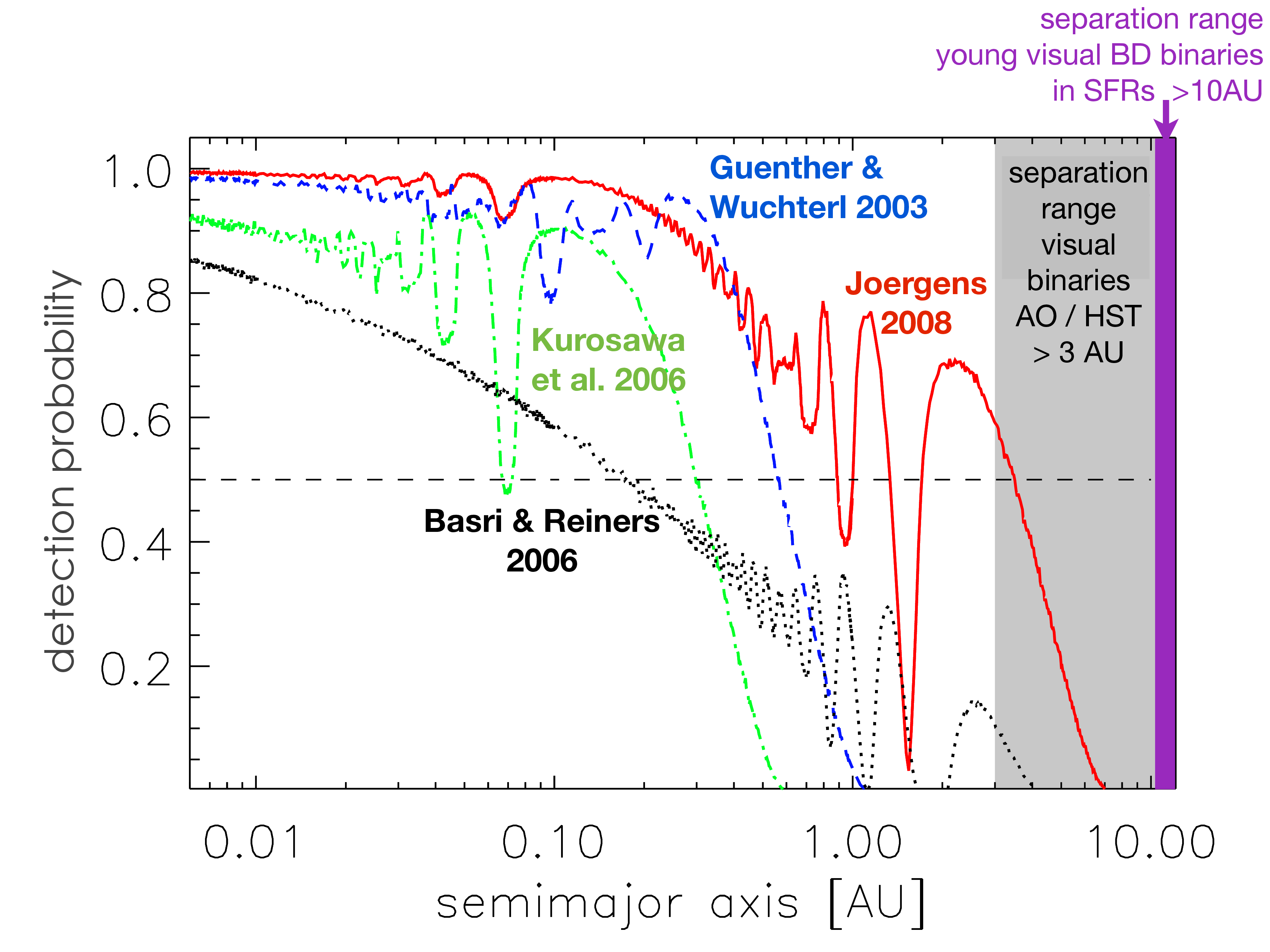}} 
   \caption{Detection probability of several RV surveys of brown dwarfs
based on Monte-Carlo simulations assuming 
a random mass ratio $M_2/M_1$ between 0.2 and 1, random orientation, 
3.3 sigma detection, and taking into account the time sampling and errors of the 
real observations. As discussed in Joergens (2008).
}
   \label{fig:pdetect}
\end{figure}
\begin{figure}[t]
  \centering
\resizebox{\hsize}{!}{\includegraphics[clip=true]{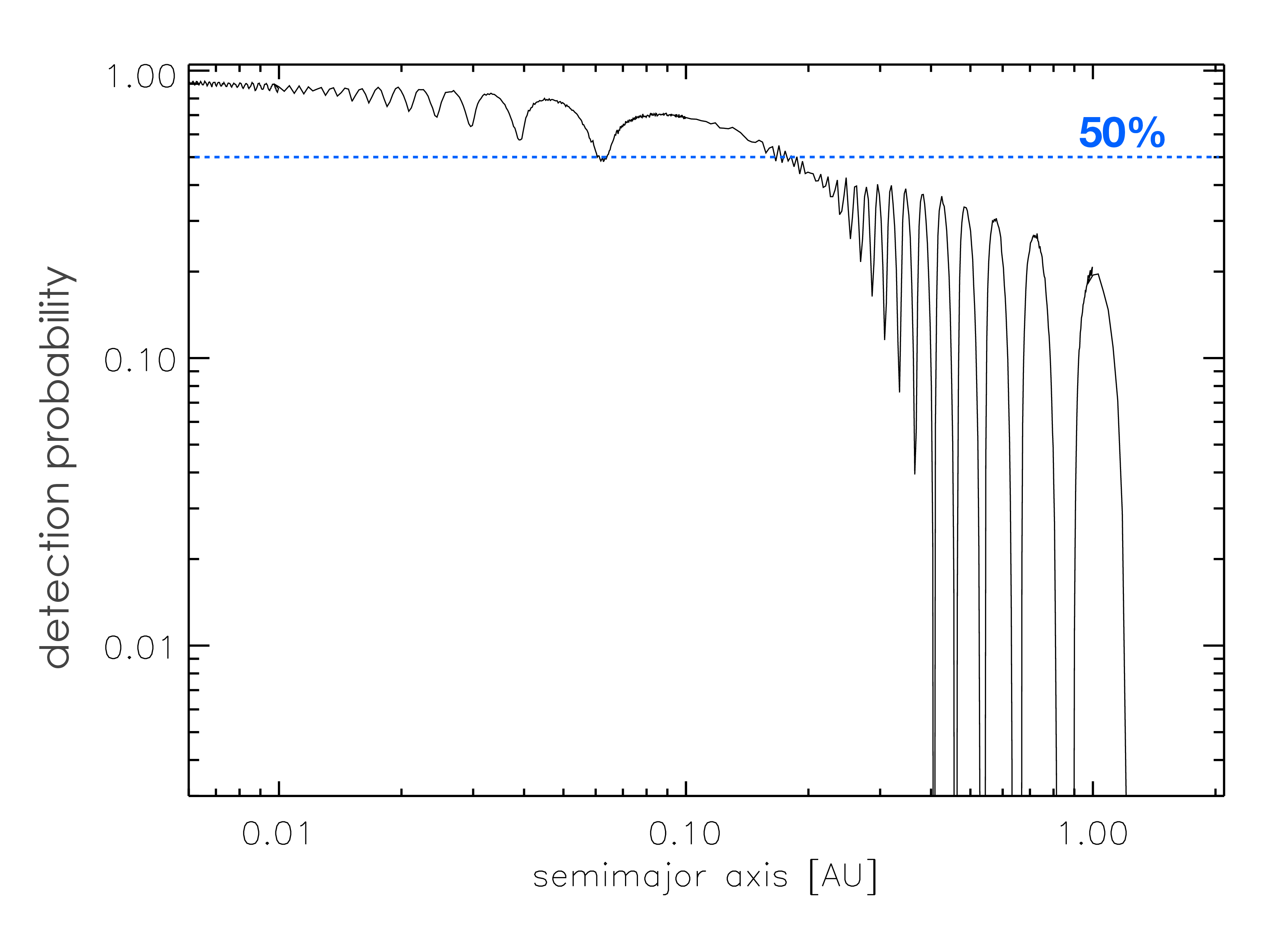}} 
   \caption{
Same as Fig.\,\ref{fig:pdetect} for the RV survey of Joergens (2008) and planet mass ratios
between 0.01 and 0.2.
}
   \label{fig:pdetectpl}
\end{figure}

The separation distribution of known brown dwarf and very low-mass binaries
(Burgasser et al. 2007)
has a peak at 3 to 10\,AU. This function is well sampled for separations 
$\gtrsim$ 3\,AU for nearby field brown dwarfs (e.g., Reid et al. 2008) and 
$\gtrsim$ 10\,AU for brown dwarfs in star-forming regions (e.g., Biller et al. 2011), as these separations 
can be directly resolved 
by Hubble Space Telescope or ground-based adaptive optics observations. 
The detection and characterization of brown dwarf binaries at close separations 
was first explored by radial velocity (RV) surveys
(e.g., Joergens 2008; Blake et al. 2010)
and recently also by ground-based astrometry 
(\citealt{sahlmann14}).  
The Gaia astrometric space mission is expected to make a significant contribution 
to the study of brown dwarf binaries at separations of a few AU, including planets orbiting brown dwarfs. 
Gaia will probe the frequency of brown dwarf binaries and 
provide astrometric orbits for those brown dwarfs that are revealed during the 
data analysis process as "non-single stars" (Gaia terminology for binary systems).

There are promising indications that brown dwarfs can have planets.
Doppler surveys found in recent years that planets orbiting cool M-stars
are frequent 
(e.g. Bonfils et al. 2013; Dressing \& Charbonneau 2013).
Brown dwarfs have the basic ingredients for planet formation, 
as they are surrounded by disk material of a few Earth masses up to one Jupiter mass
(e.g., Harvey et al. 2012; Joergens et al 2012a),
and show grain growth in their disks (e.g. Ricci et al. 2013). 
Recently a free-floating planetary mass object was found to harbor a substantial
disk of more than 10 Earth masses (Joergens et al. 2013) opening the possibility of
the existence of satellites / planets around free-floating planets.
The first brown dwarfs were found to have low-mass substellar companions at a few AU
orbital distance (Joergens \& M\"uller 2007; \citealt{sahlmann13})
and even low-mass planetary companions (Bennett et al. 2008).
However, the frequency of planets around brown dwarfs is completely
unknown and might be determined by Gaia.

\section{Spectroscopic brown dwarf binaries} 

\begin{figure}[t]
  \centering
\resizebox{\hsize}{!}{\includegraphics[clip=true]{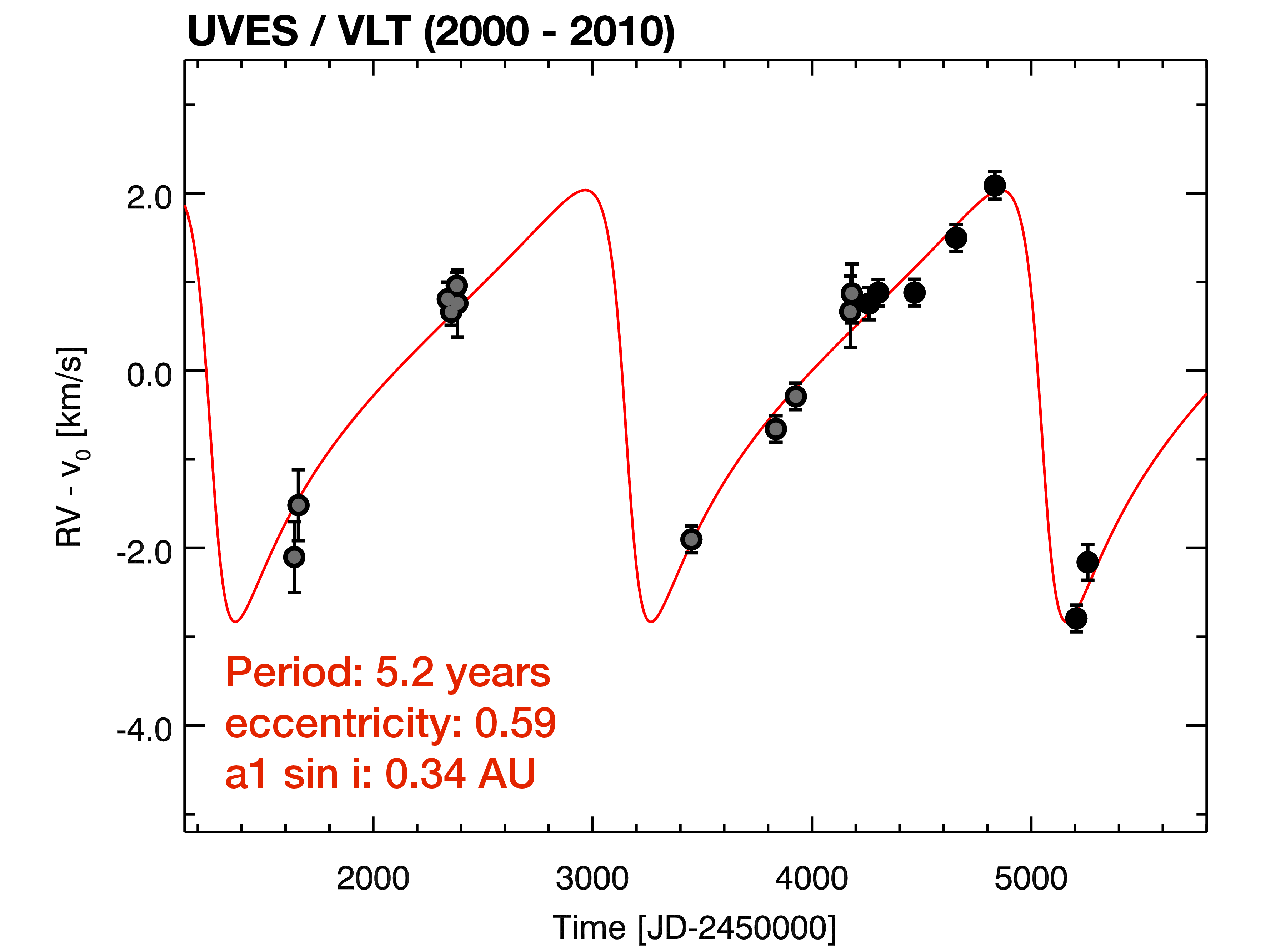}} 
   \caption{Radial velocity orbit of the young brown dwarf binary Cha\,H$\alpha$\,8 (M6) based on 
VLT/UVES monitoring. 
The best-fit Kepler orbit is shown in red. See Joergens et al. (2010).
}
   \label{fig:cha8}
\end{figure}
\begin{figure}[t]
  \centering
\resizebox{\hsize}{!}{\includegraphics[clip=true]{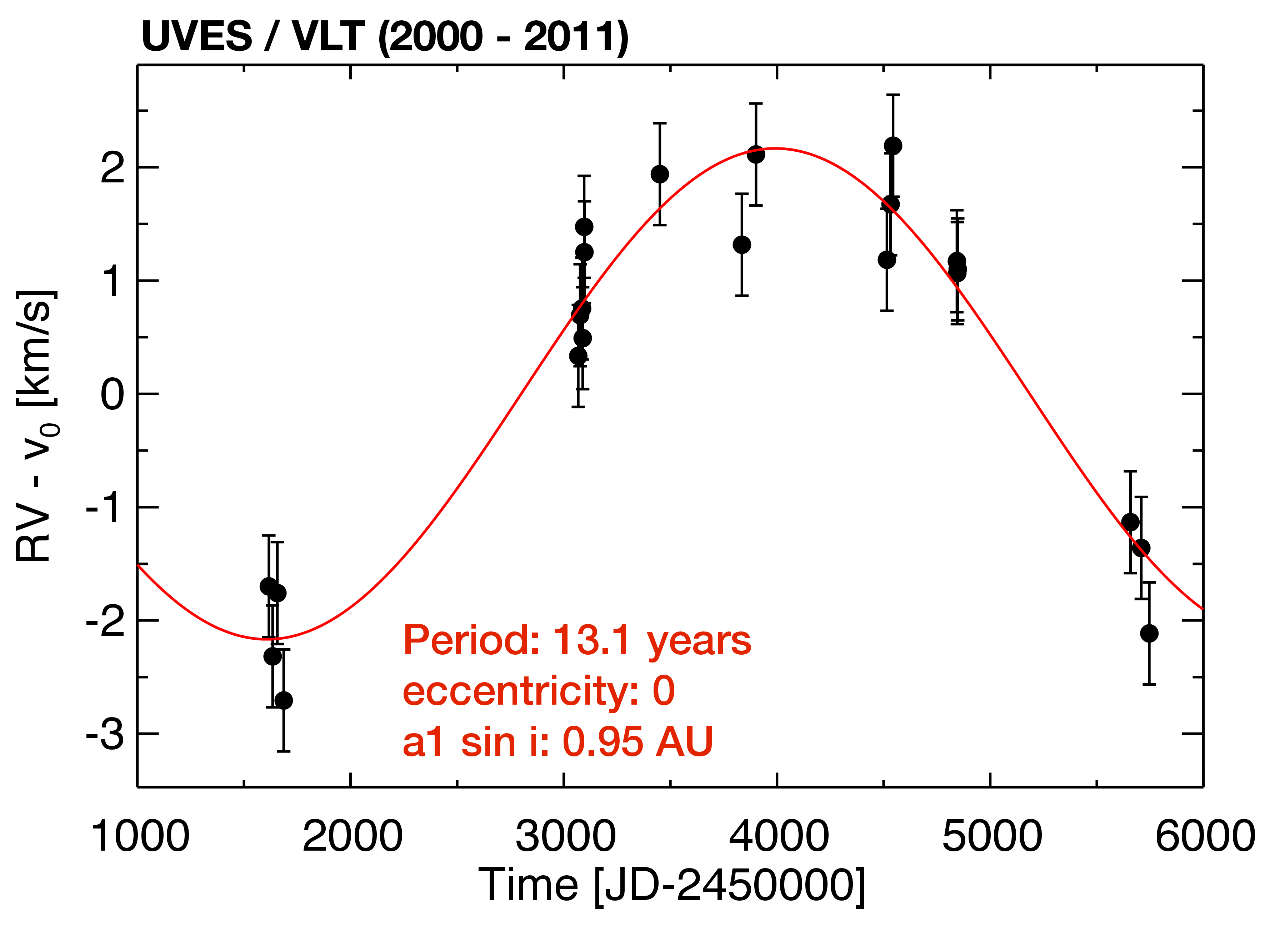}} 
   \caption{Radial velocity orbit of the young very low-mass star CHXR\,74 (M4.3) based on 
VLT/UVES data. The best-fit Kepler model is shown in red. See Joergens et al. (2012b).
}
   \label{fig:chxr74}
\end{figure}

Although one of the first detected brown dwarfs turned out to be a spectroscopic binary
(Basri \& Mart\'in 1999), the search for spectroscopic brown dwarf binaries and the determination
of their orbits remained a difficult and expensive business as it requires long-term
monitoring at large telescopes, such as the Very Large Telescope (VLT) or Keck.
As a result the number of known spectroscopic brown dwarf binaries is 
still small, with a handful of confirmed systems  (e.g., Simon et al. 2006; 
Stassun et al. 2006; Joergens \& M\"uller 2007; Blake et al. 2008).
Furthermore, most brown dwarf RV surveys probe separation ranges 
of only a fraction of an AU (cf. Fig.\,\ref{fig:pdetect}).
Exceptions are an RV survey at the VLT with UVES 
(Joergens 2008) in the young 
star forming region Chamaeleon~I (Cha\,I, 2\,Myr, 160\,pc)
that probes separation up to 3\,AU (red curve in Fig.\,\ref{fig:pdetect})
and one at the Keck telescope with NIRSPEC of field brown dwarfs 
that probes separations up to 1\,AU (Blake et al. 2010).
Both of these surveys have the sensitivity to detect RV planets in close orbits around brown dwarfs. 
Fig.\,\ref{fig:pdetectpl} displays the detection probability for planets of the RV survey in Cha\,I.
Known brown dwarf and very low-mass binary systems with orbital constraints
from RV surveys play an important role for the characterization of low-mass
binaries with Gaia because the combination of RV data and astrometry might
considerably improve the final orbit (Neveu et al. 2012).
Fig.\,\ref{fig:cha8}-\ref{fig:chxr74} display RV orbit determinations 
of two brown dwarf / very low-mass binaries (Joergens \& M\"uller 2007, Joergens et al. 2010, 2012b)
that are very promising candidates for detecting their astrometric orbits with Gaia, 
as described in the following sections.

\section{Astrometric signal of a binary}

\begin{figure*}[t]
  \centering
\resizebox{0.65\hsize}{!}{\includegraphics[clip=true]{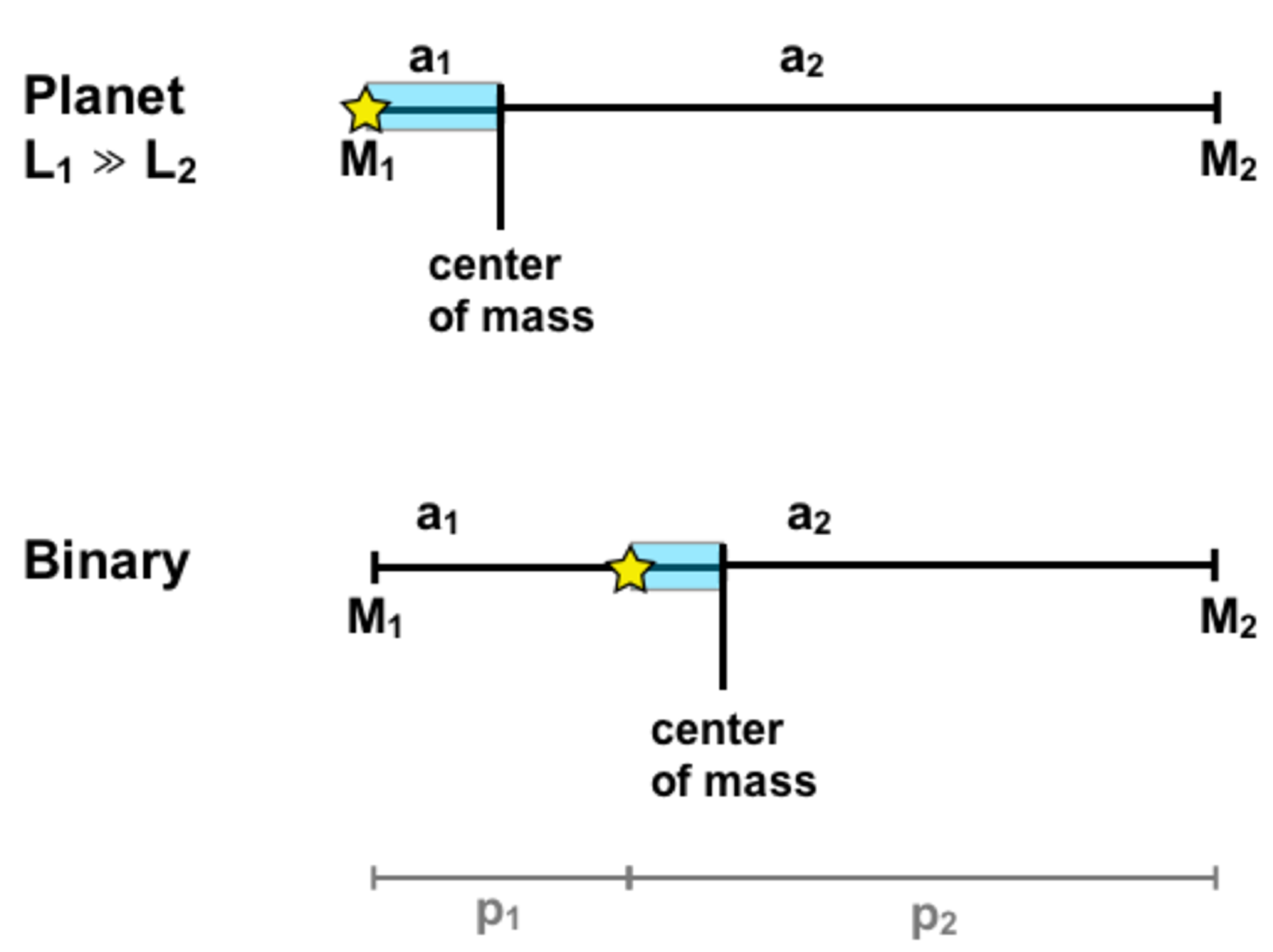} } 
   \caption{Illustration of the astrometric signal of a planet (top) and 
a binary in which the luminosity of the secondary is not negligible (bottom).
Indicated are the positions of the two components 
($M_1$, $M_2$), the center of mass, the photocenter (yellow star),
the semi-major axes of the primary's and secondary's orbit ($a_1$, $a_2$),
and the size of the astrometric displacement (blue-filled area).}
   \label{fig:astrometry}
\end{figure*}

The astrometric orbit of an unresolved binary 
is the orbit of the photocenter of the system around
the center of mass projected into the tangential plane.
For a planetary system for which the luminosity of the planet is negligible compared to that of the host star,
the astrometric orbit can be approximated by that of the primary.
This is not the case for higher mass companions, as described below.
The projection of the true orbit into the tangential plane depends on 
the inclination $i$ of the orbital plane;
$i$ remains unconstrained by RV orbit solution.

The relation for the astrometric signal in arcseconds
is given by the observable spatial displacement $dx$ at the distance $D$ (Eq.\,1).
In the case of negligible companion luminosity, 
the peak-to-peak displacement $dx$ can be expressed by twice the apparent angular size of the 
semi-major axis of the primary (Eq.\,2, Fig.\,\ref{fig:astrometry}).
It is noted that the semi-major axis of the apparent orbit is in general smaller than the
semi-major axis of the true orbit. In the most unfavorable case for astrometry 
($\omega$=90\,$^{\circ}$, $i$=90\,$^{\circ}$) 
it is identical to the semi-minor axis of the true orbit.
Using $a_1 \, M_1 = a_2 \, M_2$ one obtains the well-known relation (Eq.\,3) for the 
peak-to-peak astrometric signal
of a planet.

For binaries in which the secondary contributes significantly to the total luminosity of the system,
the photocenter is shifted from the position of the primary towards the secondary.
This is illustrated in Fig.\,\ref{fig:astrometry}. 
While a more massive companion causes a larger orbit of the primary ($a_1$),
the observable displacement decreases with larger companion mass 
because the shift of the photocenter towards the center of mass leads to a reduced observable orbit 
$a_1 - p_1$ (Eq.\,4).
We derive Eq.\,5 for the peak-to-peak 
displacement of a binary by applying $p_1 \, L_1 = p_2 \, L_2$ and $a_1 \, M_1 = a_2 \, M_2$.
One can see that 
in the limiting case of an equal-mass and equal-luminosity binary ($M_1 = M_2, L_1 = L_2$), 
the astrometric displacement $dx$ is zero.
The peak-to-peak astrometric signal for a binary 
in which the luminosity of the secondary cannot be neglected
is given in Eq.\,6.
These relations are applied in the following section to estimate the 
astrometric signal of two brown dwarf / very low-mass spectroscopic binaries.
\begin{eqnarray}
\textbf{ Planet:} \quad \quad \Theta [^{\prime \prime}] = \frac{dx \, \rm{[AU]}}{D \, \rm{[pc]}} \\
dx = 2 \, a_1 \\
\Theta [^{\prime \prime}] = \frac{2 \, a_1 \, \rm{[AU]}}{D \, \rm{[pc]}}= \frac{M_2}{M_1} \cdot \frac{2 \, a_2 \, \rm{[AU]}}{D \, \rm{[pc]}} 
\end{eqnarray}
\begin{eqnarray}
\textbf{ Binary:} \quad \quad \quad \quad dx = 2 \, (a_1 - p_1) \\
dx = 2 \left(\frac{M_2}{M_1 + M_2} - \frac{L_2}{L_1 + L_2} \right) \cdot a \\
\Theta [^{\prime \prime}] = \left( \frac{M_2}{M_1 + M_2} - \frac{L_2}{L_1 + L_2} \right) \cdot \frac{2 \, a \, \rm{[AU]}}{D \, \rm{[pc]}}
\label{equ:bin}
\end{eqnarray}

\section{Gaia astrometric orbit predictions}

\begin{figure*}
  \centering
\resizebox{\hsize}{!}{\includegraphics[clip=true]{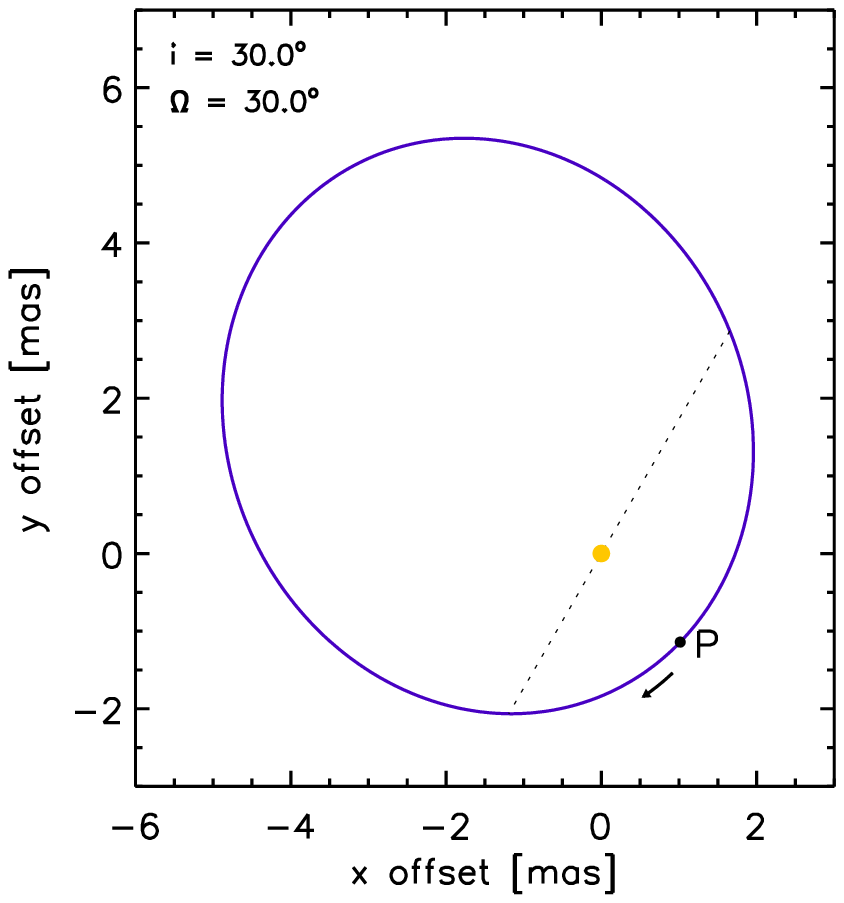}
\includegraphics[clip=true]{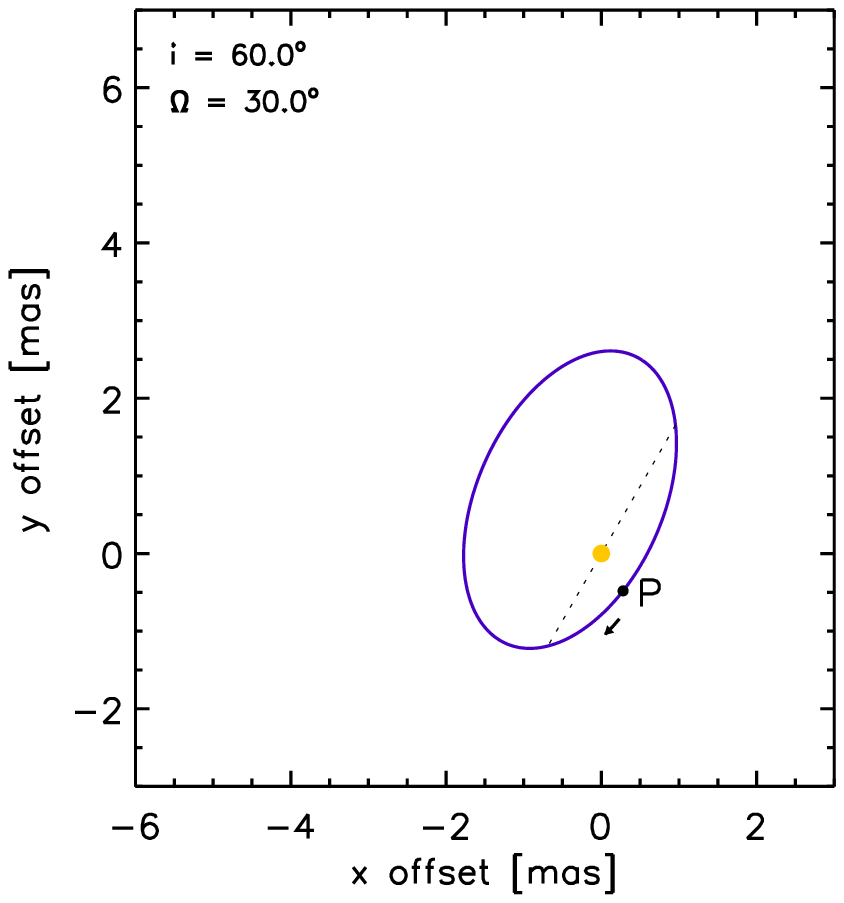}}
   \caption{
Prediction of the orbit of the primary in Cha\,H$\alpha$\,8 based on RV orbit parameters
and 
assumptions for the longitude of ascending node $\Omega$ 
and the inclination $i$. 
Note that the observable astrometric orbit is smaller than shown here because of the 
significant luminosity of the companion 
(see text).
Marked are: focal point (yellow), line of nodes (dotted line),
periastron time (P), direction of motion (arrow).
The orientation of the apparent orbit
changes as function of $i$, because varying fractions of the true
semi-major axis are hidden due to projection.
In contrast to $i$, the choice of $\Omega$ does only 
rotate the orbit in the tangential plane but
not affect its size or shape.
}
   \label{fig:astrorbit}
\end{figure*}

\subsection{Cha\,H$\alpha$\,8}
Cha\,H$\alpha$\,8 (M6) was discovered to be  
a very young spectroscopic brown dwarf binary in Cha\,I (Joergens \& M\"uller 2007).
The RV orbit solution based on 10 years of UVES monitoring 
with high-spectral resolution at the VLT (Fig.\,\ref{fig:cha8}) has an 
orbital period of 5.2\,years, an eccentricity of 0.59, and 
a semi-major axis of the primary of $a_1 \sin i=0.34$\,AU (Joergens et al. 2010).

With a $V$ magnitude of 20.1\,mag (Comer\'on et al. 2000) Cha\,H$\alpha$\,8 should 
be easily
observable with Gaia.
Gaia's faint star limit of $G\sim20$\,mag (http://sci.esa.int/gaia)
translates for brown dwarfs to a limit in $V$ of about 23\,mag or even slightly fainter.
This follows from 
using $(V-I_C)\approx4.30$ from Kenyon \& Hartmann (1995) and the relation
to convert $V$ to $G$ magnitudes given $(V-I_C)$ from Jordi et al. (2010),
which yields a $(V-G)$ color of an M6 dwarf of about 3\,mag, and even larger for later-type brown dwarfs.

To estimate the peak-to-peak astrometric signature of Cha\,H$\alpha$\,8 
we first consider the size of the apparent orbit of the primary.
The minimum apparent angular size of twice the semi-major axis of the primary
is given by twice the \emph{semi-minor}
axis of the true orbit of the primary 
for an inclination of 90\,$^{\circ}$. Using the determined RV orbit parameters 
yields a minimum positional displacement of
0.54\,AU, corresponding to 3.4\,mas at a distance of 160\,pc.
For smaller inclinations, the astrometric orbit would be significantly larger.
We performed simulations of the orbit of the primary around the center of mass 
for Cha\,H$\alpha$\,8 (cf. Fig.\,\ref{fig:astrorbit}) and find 
as apparent size of the primary's orbit about 4\,mas for $i$=60\,$^{\circ}$
and about 7\,mas for $i$=30\,$^{\circ}$, resp.

To derive the astrometric signal of Cha\,H$\alpha$\,8 the 
luminosity of the
companion has to be taken into account.
We estimate the luminosity ratio of the two components of Cha\,H$\alpha$\,8 as $L_2/L_1 \geq 0.18$ 
based on the RV orbit and evolutionary models (Baraffe et al. 1998). 
By applying Eq.\,\ref{equ:bin} we predict a peak-to-peak astrometric signature 
of Cha\,H$\alpha$\,8 of $1.2-1.6$\,mas for $i=60^{\circ}-90^{\circ}$.
This should be detectable with Gaia given the envisioned performance of this space mission.

\subsection{CHXR\,74}

The young very low-mass star CHXR\,74 (M4.3) was detected to be
a spectroscopic binary in Cha\,I based on long-term UVES monitoring (Joergens et al. 2012b). 
The best-fit RV orbit (Fig.\,\ref{fig:chxr74})
has a period of 13.1\,years, $e=0$, and $a_1 \sin i=0.95$\,AU. From non-detection of the long-period companion 
in high-resolution NACO/VLT images, an upper limit of the companion 
mass and a minimum inclination angle of 40$^{\circ}$ were derived (Joergens et al. 2012b).
Considering different inclinations between 90$^{\circ}$ and 40$^{\circ}$ and taking 
the luminosity of the companion into account ($L_2/L_1=0.3-0.5$), we predict  
a peak-to-peak astrometric signal between 0.4 and 0.8\,mas for CHXR\,74.
This should be detectable with Gaia for this V=17.3\,mag object.
While the duration of the Gaia mission is shorter than the orbit of CHXR\,74,
combining RV data and RV orbital parameters with
Gaia astrometric data will likely allow the determination of 
its astrometric orbit.

\section{Conclusions}

We presented predictions of the astrometric signal of 
the two brown dwarf / very low-mass spectroscopic binaries Cha\,H$\alpha$\,8 
and CHXR\,74 being of the order of 0.5\,mas to more than 1\,mas.
These astrometric orbits should be easily detectable with Gaia.
The combination of their RV data from our survey and Gaia astrometry might
considerably improve their final orbits.
Both systems can play a key role because they are very young and provide constraints 
for the early evolution of very low-mass objects. Furthermore, they are two representatives 
of only a handful of brown dwarf 
and very low-mass RV binary systems with solved orbits. Their relatively long orbital 
periods ($>$5\,yrs) qualify them as promising Gaia targets.

\bibliographystyle{aa}

\end{document}